\documentclass[sigconf]{acmart}
\setcopyright{none}
\renewcommand\footnotetextcopyrightpermission[1]{}
\acmConference{}{}{}
\acmBooktitle{}
\acmPrice{}
\acmISBN{}
\acmDOI{}
\AtBeginDocument{%
  }

\usepackage{subcaption}
\usepackage{float}
\usepackage{array}
\usepackage{caption}
\usepackage{subcaption}
\usepackage{tabularx}
\usepackage{multirow} 
\usepackage{enumitem}
\usepackage{geometry}
\usepackage{longtable}
\usepackage{footmisc}
\usepackage{booktabs}
\usepackage{threeparttable}
\usepackage{pifont}  
\usepackage[font=small,labelfont=bf]{caption} 
\usepackage{booktabs}
\usepackage[table]{xcolor}
\usepackage{graphicx}
\usepackage{ragged2e}
\usepackage{stfloats}
\usepackage{makecell}
 
\definecolor{LowColor}{RGB}{198, 239, 206}     
\definecolor{ModColor}{RGB}{255, 235, 156}     
\definecolor{HighColor}{RGB}{255, 199, 206}

\title{Designing Computerized Gait Analysis for Pediatric Care: Clinician Perspectives on Sensing, Workflow, and Care Environments}

\author{Elizabeth Hong}
\authornote{These authors contributed equally to this work.}
\affiliation{%
  \department{Computer Science}
  \institution{Stanford University}
  \country{USA}
}

\author{Andrea Green}
\authornotemark[1]
\affiliation{%
  \department{Civil and Environmental Engineering}
  \institution{Stanford University}
  \country{USA}
}

\author{Ge Wang}
\orcid{0000-0003-4656-2175}
\affiliation{%
  \department{Siebel School of Computing and Data Science}
  \institution{University of Illinois Urbana-Champaign}
  \country{USA}
}

\author{Yiwen Dong}
\orcid{0000-0002-7877-1783}
\affiliation{%
  \department{Industrial \& Enterprise Systems Engineering}
  \institution{University of Illinois Urbana-Champaign}
  \country{USA}
}

\begin{document}

\begin{abstract}
Computerized gait analysis (CGA) is a diagnostic tool for various disorders in children, enabling objective assessment of movement for clinical interventions. Existing work on pediatric CGA has focused on technical/clinical performance, leaving less understood about how clinicians interact with these systems in practice. We interviewed 12 pediatric clinicians and one system designer experienced with CGA. Participants described mismatches between CGA and pediatric care, including children's sensory sensitivities to wearables, difficulties placing sensors on child-sized bodies, and challenges with engagement during data collection. These mismatches required clinicians to adapt procedures to children's bodies, behaviors, and needs. Clinicians highlighted opportunities to extend CGA into environments such as playgrounds, where children’s movement could be assessed in contexts relevant to their development. Based on these insights, we offer design implications for pediatric CGA, including designing sensing and calibration for smaller and changing bodies. Our study informs clinical technologies that account for pediatric needs.
\end{abstract}


\maketitle

\section{Introduction}
Gait analysis is widely used in clinical evaluation across neurological, musculoskeletal, and cardiovascular medicine, playing a critical role in managing conditions ranging from cognitive disorders to pediatric neuromuscular and neurological conditions such as muscular dystrophy and cerebral palsy~\cite{jayakody2019gait,verghese2007quantitative,seals2022they,dong2024ambient,dong2020md,armand2016gait}. Gait, or a person's walking pattern, can reflect underlying neuromuscular function, joint function, and disease progression, providing clinicians with important information for diagnosis and prognosis~\cite{hulleck2022present}. Deriving these clinical insights requires capturing and evaluating characteristics of a person's gait. Historically, gait analysis relied on visual observation by trained clinicians~\cite{hulleck2022present,wren2011efficacy}. However, visual assessments can be subjective and limited in the range of gait characteristics clinicians can observe and quantify~\cite{hulleck2022present,wren2011efficacy}. Computerized gait analysis (CGA) introduces sensors and computational infrastructure to provide more objective and comprehensive gait measurements~\cite{shrader2021instrumented,vun2024vision,dong2024ambient,mason2023wearables}, showing accuracy comparable to gold-standard tools~\cite{jagos2017mobile}.

CGA utilizes vision-based systems, ambient sensors, and wearable devices to collect three-dimensional measurements, objective neuromuscular profiling, and longitudinal tracking that exceed what visual observation alone can provide~\cite{seals2022they}. These computerized methods are particularly important for pediatric patients, where comprehensive gait analysis and timely intervention can substantially improve long-term mobility and quality of life~\cite{dumuids2022effects,shrader2021instrumented}. Yet, these capabilities also make gait measurement dependent on interactions among patients' bodies, sensors, clinicians, and the environments in which measurement occurs. The quality of computationally derived gait data depends not only on sensing and algorithmic performance but also on whether patients can participate in measurement in the ways these systems expect~\cite{gagnon2012systematic, lee2024physical}. Existing work has identified barriers to integrating CGA into clinical workflows, including lengthy setup times, specialized technical expertise, and difficulties interpreting large volumes of data~\cite{hebda2022clinicians}. 

Less attention, however, has been paid to the interaction work required to make CGA possible when patients do not conform to the assumptions around which sensing systems and assessment procedures were designed. This factor presents particular challenges in pediatric care, as assessing children requires clinicians to consider their developmental level, ability to participate in an assessment, and caregiving context~\cite{srinath2019clinical}. Pediatric assessments may also require clinicians to adjust how they engage with children and involve parents and caregivers to support their participation~\cite{srinath2019clinical}. However, this work is largely absent from technical accounts of CGA systems, yet it is central to whether assessments succeed and produce accurate data~\cite{obrien2021not, sivaraman2023ignore}. Pediatric CGA therefore provides a useful case for examining a broader HCI challenge: how should computational sensing systems accommodate people whose bodies, behaviors, and contexts do not align with standardized system assumptions? Examining pediatric care allows us to understand not only where these assumptions break down but also how clinicians compensate for them in practice and how interaction design might reduce rather than reproduce this burden.

To investigate these questions, we conducted a qualitative interview study supplemented by a structured survey with 13 experts who regularly engage with CGA in pediatric settings: 12 pediatric clinicians and one system designer. We examine CGA not only as a clinical measurement technology but also as a sociotechnical system in which sensing, children, clinicians, families, and care settings collectively shape how gait data are produced and interpreted. The pediatric clinicians in our study provide healthcare to infants, children, adolescents, and young adults across various neuromuscular and neurological conditions. The system designer contributed expertise in developing CGA systems across multiple use cases. We further categorize clinicians into two groups based on their roles in the gait analysis process: medical experts ($n$=7), who typically lead the analysis by coordinating technical procedures and facilitating patient interactions, and physicians ($n$=5), who interpret the resulting data to inform patients and their families about diagnoses and treatment plans.

Our study explored the following research questions:

\begin{enumerate}[leftmargin=5em]
    \item[\textbf{RQ1:}] How do pediatric clinicians perceive, adapt, interpret, and negotiate computerized gait analysis when working with child patients?

    \item[\textbf{RQ2:}] What design requirements emerge for pediatric gait analysis systems that align sensing, clinical workflows, and care settings?
\end{enumerate}

We use \textit{perceive} to refer to how clinicians evaluate CGA; \textit{adapt} to describe how they modify their practices or use of CGA to accommodate child patients; \textit{interpret} to refer to how they make sense of computationally derived gait data in relation to their clinical judgment; and \textit{negotiate} to describe how they manage tensions between the requirements of CGA and the needs of patients, clinical workflows, and care settings.

\section{Contribution}
With this work, we make the following contributions: 

\begin{itemize}
    \item Child-specific mismatches and clinician workarounds in pediatric CGA: Mismatches between CGA systems and children's needs, including challenges placing markers on small and variable bodies and managing the sensory burden of sensors.

    \item Design guidelines for pediatric CGA: Designing CGA to better support pediatric care, including calibration that accommodates smaller and changing bodies, sensing approaches that reduce physical contact when possible, and sensing that accounts for the environment in which gait is measured.

    \item Design implications for designing pediatric healthcare technology: When sensing systems do not account for differences in development or sensory needs, clinicians may take on additional work to make these systems usable in practice. Pediatric gait assessment highlights a broader HCI challenge of designing technologies for populations whose bodies, behaviors, and care settings may not align with the assumptions embedded in these systems.
\end{itemize}

\section{Related Work}
In this section, we present related work on computerized gait analysis and sensing in pediatric care, designing health technologies for children, and interaction work with clinical technologies. 

\subsection{Computerized Gait Analysis and Sensing in Pediatric Care}
Computerized gait analysis (CGA) allows clinicians to objectively and comprehensively assess how patients walk and use this information to support diagnosis and treatment of various neurological, musculoskeletal, and cardiovascular conditions. CGA can reveal movement patterns associated with pediatric conditions such as cerebral palsy and muscular dystrophy, including changes in joint movement and reduced walking speed~\cite{armand2016gait,carcreff2018best}. Pediatric CGA is valuable for understanding complex gait patterns and supporting treatment decisions, particularly for children with cerebral palsy~\cite{armand2016gait,shrader2021instrumented}. This is particularly important for pediatric care, where early identification and intervention can substantially improve patient outcomes~\cite{dumuids2022effects,wren2009effects}. Computerized gait analysis combines sensing hardware with software and algorithms to derive quantitative spatiotemporal, kinematic, and kinetic measures of walking, including gait speed, step and stride length, joint angles, and ground reaction forces~\cite{muro2014gait,minea2025using}. These measures support clinicians' interpretation and identification of movement problems and inform decisions about surgery, therapy, and rehabilitation~\cite{hulleck2022present,shrader2021instrumented}. 

Prior work has identified barriers to utilizing CGA in practice, including cost, limited access, specialized training, and difficulties integrating gait analysis into routine clinical practice~\cite{hebda2022clinicians}. Additionally, the reliance on specialized laboratory equipment and controlled testing procedures may alter a child's natural gait~\cite{bourgeois2014spatio}. Current research has explored more practical, portable, and flexible sensing approaches that could extend gait assessment beyond the laboratory. Wearable sensors, for example, have been evaluated as alternatives for measuring children's gait outside constrained laboratory environments~\cite{bourgeois2014spatio,carcreff2018best,van2013gait}, while newer work has explored markerless and ambient sensing approaches~\cite{wishaupt2024applicability,dong2024ambient}. Across this work, a central concern has been whether emerging technologies can accurately and reliably capture gait measures in children. Yet comparatively less attention has been given to what happens during the assessment itself and how clinicians work with children to position, wear, and use these sensing technologies or adapt the process when standardized procedures do not fit the child.

\subsection{Designing Health Technologies for Children}
Prior HCI research has highlighted the need to design health technologies around the particular needs of children and pediatric care. Pediatric health technologies must account for developmental and behavioral differences that distinguish interactions with children~\cite{srinath2019clinical}. Children's ability and willingness to participate in clinical assessments can vary with their developmental level, emotional state, and comfort, requiring clinicians to adapt how assessments are conducted and how children are engaged~\cite{srinath2019clinical}. Pediatric assessment requires clinicians to build rapport through familiarity with children's interests, such as media, games, and toys~\cite{srinath2019clinical}. However, how clinicians perform this interaction work while simultaneously managing the technical demands of pediatric CGA remains underexplored.

Pediatric assessment also extends beyond the individual child, often requiring information and involvement from parents, caregivers, teachers, and clinicians across different settings~\cite{srinath2019clinical,sepehri2023binder}. As children grow, collaborative health-tracking technologies must also balance children's independence with the involvement of parents and clinicians~\cite{cha2025collaborative}. Further, children's experiences in clinical settings can be shaped by environmental context, with features such as lighting and color impacting anxiety, stress, and perceived pain~\cite{robinson2015ambient,dalke2006colour}. While prior work demonstrates the importance of designing around children's experiences and care relationships, less is known about how children's needs and care settings shape clinicians' interactions with sensing technologies during pediatric CGA; therefore, we examine this intersection in this work. 

\subsection{Interaction Work with Clinical Technologies}
HCI research has long examined the work people perform to make technologies function in practice. We use \textit{interaction work} to describe the practical work required to make interactions between people and technologies function in real settings. This framing builds on related concepts such as articulation work, which describes the work required to coordinate tasks, people, resources, and timing~\cite{star1999layers,huber2023navigating}; invisible work, which captures necessary but often unrecognized work~\cite{stisen2016accounting,star1999layers}; and workarounds, where people adapt technologies that do not fully fit their needs~\cite{wibisono2019understanding}. These concepts show that when technologies do not account for everyday practice, the work required to make them function can shift onto users. This issue is particularly important in healthcare. Clinicians may need to adapt technologies based on environmental conditions, workflows, and clinical judgment~\cite{beede2020human,sivaraman2023ignore,ng2019provider}. Interaction work also involves patients, as clinicians rely on contextual information, such as pain and fatigue, for decision-making that sensing technologies may not capture~\cite{akinsiku2021not,obrien2021not}. Thus, making a clinical technology work involves interactions among the technology, patients, and the clinical environment.

Within CGA, clinicians have identified data complexity, limited normative reference data, and poor workflow integration as barriers to using quantitative gait sensing~\cite{obrien2021not,hebda2022clinicians}. How gait data are presented can also shape whether clinicians can interpret and use the resulting information~\cite{seals2022they}. However, this work has primarily examined how clinicians interact with sensing systems and their outputs, rather than how interactions with patients and care settings shape the use of these technologies. Research on wearable diagnostic systems for children shows that sensor-based assessment involves fitting and placing multiple sensors on children's bodies~\cite{jiang2020weda}, demonstrating that measurement itself requires interaction with patients. Additionally, the existence of guidelines specifically for assessing children reflects a need for clinical practices tailored to pediatric assessment~\cite{srinath2019clinical}. Introducing computational and sensing technologies into pediatric assessment therefore raises questions about how the requirements of these technologies align with existing assessment practices and situated care settings.

\section{Method}
We conducted an IRB-approved study (protocol ID anonymized) with pediatric clinicians and a designer of computerized gait analysis (CGA) systems. In this section, we introduce our participants, the study procedure, and the data analysis method.  

\subsection{Participants}
Between June and October 2025, we conducted 13 one-on-one sessions (survey and interview) with seven medical experts, five physicians, and one designer. We included both medical experts and physicians to capture complementary perspectives: medical experts because of their specialized expertise in conducting gait analyses, rehabilitation planning, data collection, and biomechanical data interpretation; and physicians because of their broader clinical experience in conducting gait analyses, diagnosis, treatment planning, and integrating gait analysis findings into patient care. We included the perspective of a CGA system designer as a domain expert. Demographic and background information about participants is provided in Table~\ref{tab:surveydata}. 

\begin{table}[t]
\centering
\caption{Participant background and demographics}
\label{tab:surveydata}

\small
\setlength{\tabcolsep}{3pt}
\renewcommand{\arraystretch}{1.05}

\begin{tabularx}{\columnwidth}{@{}l l X c@{}}
\toprule
\textbf{ID} & \textbf{Education} & \textbf{Professional title} & \textbf{Years} \\
\midrule
P1-Exp  & Doctorate & Physical Therapist & 7--8 \\
P2-Exp  & Master's & Biomechanist Technical Supervisor & 3--4\\
P3-Exp  & Master's & Doctoral Candidate & 3--4\\
P4-Exp  & Bachelor's & Motion Lab Technician & $>$10\\
P5-Exp  & Doctorate & Physical Therapist Supervisor & 3--4\\
P6-Exp  & Master's & Athletic Trainer & $>$10\\
P7-Exp  & Doctorate & Physical Therapist & 3--4\\
P8-Phy  & Doctorate & Pediatric Orthopedic Surgery & 3--4\\
P9-Phy  & Doctorate & Pediatric Orthopedic Surgery & $>$10\\
P10-Phy & Doctorate & Pediatric Orthopedic Surgery & 5--6\\
P11-Phy & Doctorate & Pediatric Orthopedic Surgery & $>$10\\
P12-Phy & Doctorate & Pediatric Orthopedic Surgery & --\\
P13-Des & Doctorate & Postdoctoral Scholar & 1--2\\
\bottomrule
\end{tabularx}

\vspace{0.25em}
\raggedright
\tiny
\textbf{Abbrev.:} Exp = Medical Expert; Phy = Physician; Des = Designer.
\end{table}

\subsection{Procedure}
\subsubsection{\textbf{Recruitment}
}
Participants were recruited from a university hospital gait laboratory [anonymized for review] beginning in May 2025. Recruitment was conducted on a rolling basis via email invitations sent to all eligible clinicians affiliated with the gait lab and word-of-mouth recruitment within participants' professional network. To be eligible, clinicians were required to have (1) experience conducting computerized gait analysis and (2) direct interactions with child patients. 

\subsubsection{\textbf{One-on-one Sessions}}
At the beginning of the session, participants were introduced to the research objective and completed an online consent form, which included agreement to audio and video recording. Participants were then presented with the following scenario to contextualize the questions within a pediatric context: 

\begin{quote}
\textit{A pediatric patient is currently undergoing gait analyses to guide intervention for cerebral palsy. As part of this process, the attending clinician (medical expert or physician) will choose a configuration of sensors to perform the gait analysis. However, to ensure that the process is robust, sensitive, efficient, accurate, and comfortable in the context of analyzing pediatric patients’ gait, the expert has asked for your input to help inform their final selection.}
\end{quote}

Participants then completed a quantitative survey, followed by a semi-structured interview based on the scenario. After the interview, participants completed an online demographic survey. The surveys and interview questions were designed by the research team. All sessions were conducted via Zoom and lasted approximately one hour. 

\paragraph{\textbf{Complementary surveys}}
We conducted two complementary surveys to capture clinician perspectives on CGA. Each survey was tailored to the expertise of the two clinician groups (medical experts and physicians). Medical experts often interact with the technical and biomechanical aspects of gait analyses, so their survey focused on evaluating the performance of the CGA sensing approaches and their applicability in various contexts. Physicians evaluate the results of gait analyses and use them to inform treatment planning. Therefore, their survey focused on evaluating the importance of measuring various gait parameters for decision-making. Both surveys also asked participants to prioritize performance criteria when selecting sensing approaches. These insights provided a foundation for understanding the perceived strengths and applications of CGA. An overview of each survey can be seen in Appendix \ref{app:survey}.

\begin{figure}[H]
\centering
\includegraphics[width=0.9\linewidth]{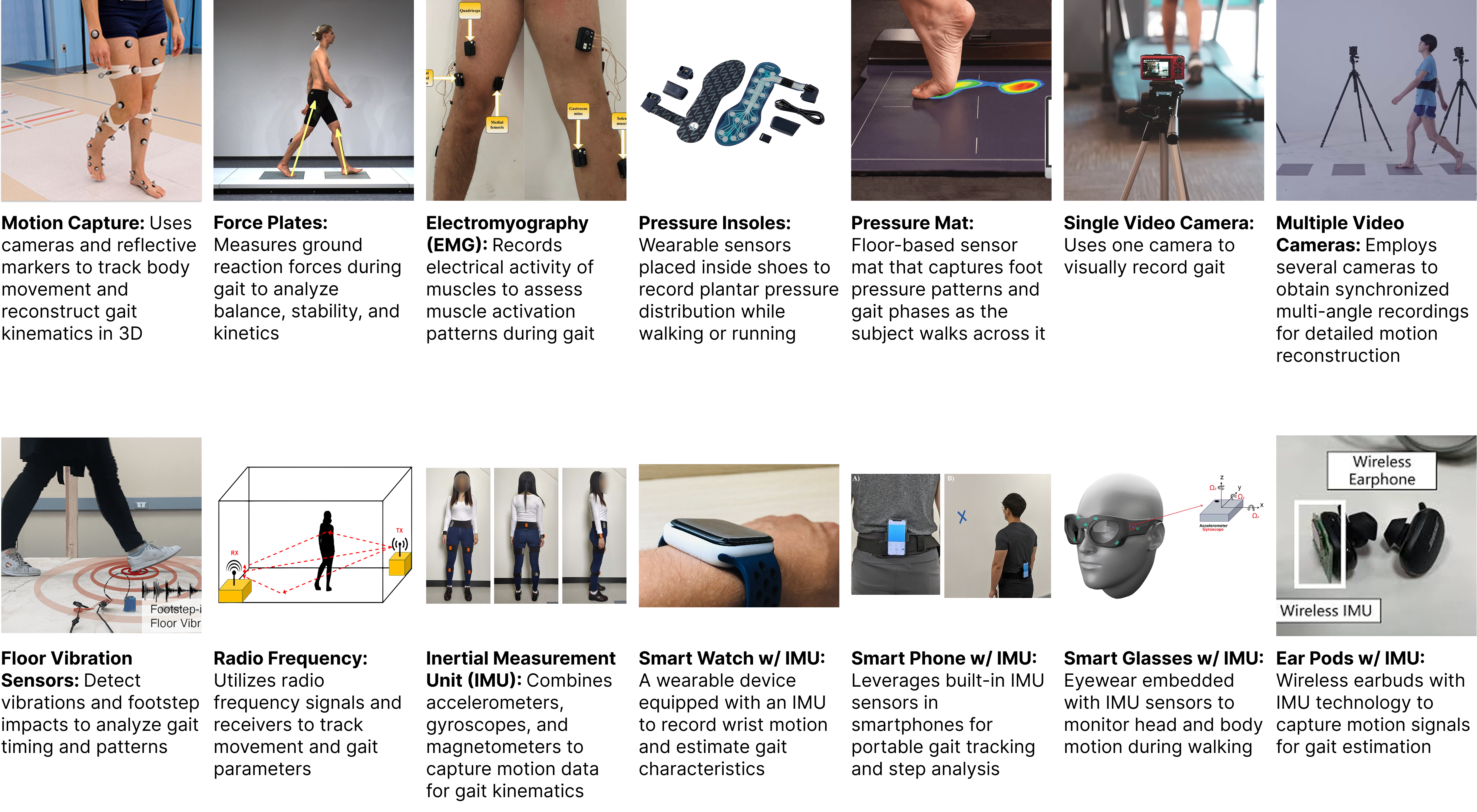}
\caption{The CGA sensing approaches that were presented to participants. Participants evaluated each approach for functionality, accuracy, usability, affordability, patient comfort, availability, training requirements, and scalability. They also selected which approaches they would ``definitely use,'' ``maybe use,'' or ``not use'' in a gait lab, local care center, and patient home.}
\label{fig:approaches}
\end{figure}

\begin{itemize}
    \item {\textbf{Medical expert survey:}}
The survey for medical experts ($n=7$) evaluated CGA sensing approaches with an emphasis on sensor functionality, accuracy, and clinical applicability. The survey was structured into four main sections. In the first section, participants ranked the overall importance of seven evaluation criteria, including functionality, accuracy, ease of use, comfort, affordability, accessibility, and scalability, using a seven-point Likert scale to align with the seven criteria (1 = not important, 7 = extremely important)~\cite{finstad2010usability}. The second section captured perceptions and prior experience with 14 CGA sensing approaches (Figure \ref{fig:approaches}). 

They rated each approach in functionality (i.e., ability to measure five gait parameters: temporal parameters, spatial parameters, joint angles/kinematics, ground reaction forces/kinetics, and muscle activation status), accuracy for each parameter, ease of use, patient comfort, affordability, availability, and scalability. Third, participants evaluated each approach for deployment feasibility in three care settings: a specialized gait lab, a local care center, and a patient’s home. These settings were selected to represent the spectrum of environments in which pediatric CGA is currently conducted or envisioned~\cite{hebda2022clinicians, gu2026advancements}. The fourth and final section invited open-text comments to provide additional feedback on CGA. 

\item{\textbf{Physician survey:}} While medical experts interact primarily with the sensor and data collection layer of the CGA pipeline, physicians engage with the output end, interpreting visualized results to inform diagnosis and treatment. Therefore, the physician survey focused on the clinical decision-making value of measuring gait parameters rather than sensor performance. Physicians evaluated temporal and spatial parameters, joint angles/kinematics, ground reaction forces/kinetics, and muscle activation status. In the first section, participants rated the importance of each parameter as low, moderate, or high for decision-making. The second section addressed the importance of these parameters across the three clinical contexts—specialized gait lab, local care center, and patient home—using the same three-level scale. The survey concluded with an open-text section allowing physicians to provide qualitative feedback regarding gait analysis parameters. For both surveys, all questions also had an option for participants to select if they were not familiar with the approach or parameter in question.
\end{itemize}

\paragraph{\textbf{Semi-structured interviews}}
All participants took part in a semi-structured interview lasting approximately 30 minutes. The interviews supplemented the survey data by providing nuanced qualitative insights. The interviews began with a short self-introduction, where participants described their role, background, and workflow in child gait analysis. The second portion focused on experiences with CGA, including comparisons with traditional manual approaches, advantages and disadvantages, and perceived impacts on clinical practice. The final portion of the interview was the ideation session, where participants were asked to envision design needs, discussing potential improvements to analysis processes, care settings, and technology design, as well as the roles of clinicians, children, and families in shaping these systems. 

\subsection{Data Analysis}
\paragraph{\textbf{Quantitative}}
Survey responses were analyzed using descriptive statistics (mean and standard deviation) to identify patterns in participants' perceptions of CGA sensing approaches (RQ1) and how they prioritized performance and gait parameters. Given the relatively small sample size, the analysis was intended to highlight observable trends and areas of consensus rather than produce generalizable inferences. The primary objective was to capture expert perspectives to reveal both points of agreement and divergence across participants.  

To evaluate judgments on the scope of different approaches, participants indicated whether each method could capture five gait parameters: temporal parameters, spatial parameters, joint angles/kinematics, ground reaction forces/kinetics, and muscle activation status. These binary (yes/no) questions were summarized using the mode response to reflect the most common participant view. Responses regarding the likelihood of adopting each approach for use in three contexts---\textit{gait laboratory, local care center, and patient home} were summarized using the top two approaches with the most ``definitely use'' responses to show which methods were viewed as most applicable in each context, highlighting if the setting influenced adoption preferences. The top two approaches were highlighted to reflect the greatest consensus and indicate those most likely to be adopted in practice.

\paragraph{\textbf{Qualitative}}
All interviews were video- and audio-recorded with participants’ consent obtained via digital forms. One researcher transcribed the recordings. Two researchers analyzed the data using a thematic approach~\cite{braun2006using}. Initially, two researchers independently took notes of 100\% of the data to develop a preliminary codebook. They then discussed individual codes, resolved disagreements, and finalized a shared codebook, with interrater reliability (IRR) exceeding 90\%. Using this final codebook, one researcher coded 100\% of the data, while an additional researcher conducted a validity and reliability check on a random sample of approximately 30\% of the total data (IRR >90\%). The finalized codebook included themes on perceptions of and experiences using CGA with child patients, including challenges, benefits, and barriers to implementation (RQ1) and participants' envisioned design needs (RQ2). The most significant themes were then presented as the findings and supported with quotes from participants. 

\subsection{Ethical Considerations}
We recognize the sensitivity of conducting research with human participants and took comprehensive precautions throughout the study. The research protocol received institutional review board approval, ensuring full compliance with ethical standards. Before data collection, all participants were informed about the study’s purpose, procedures, and their rights, including the option to withdraw at any stage without consequence. Informed consent was obtained from participants digitally. Given the clinical context, particular care was taken to safeguard any information relating to children. Study sessions conducted via Zoom were recorded solely for research purposes. To protect participant privacy, all audio and video recordings were stored on an encrypted drive accessible only to the research team. 

\section{Findings}
The findings are organized into two parts. First, we present quantitative survey results on participants' perceptions of 14 computerized gait analysis (CGA) sensing approaches (\textbf{RQ1}). Second, we present qualitative findings on how clinicians adapt, interpret, and negotiate CGA when working with child patients (\textbf{RQ1}), and the design requirements that emerge for aligning CGA with clinical workflows, pediatric patients, and care settings (\textbf{RQ2}). Across the findings, three patterns were especially important for design: clinicians value CGA for producing objective evidence; working with child patients reveals mismatches between CGA and pediatric care; and future CGA should align sensing with patients and care settings.

\begin{figure}[H]
    \centering
    \includegraphics[width=0.75\linewidth]{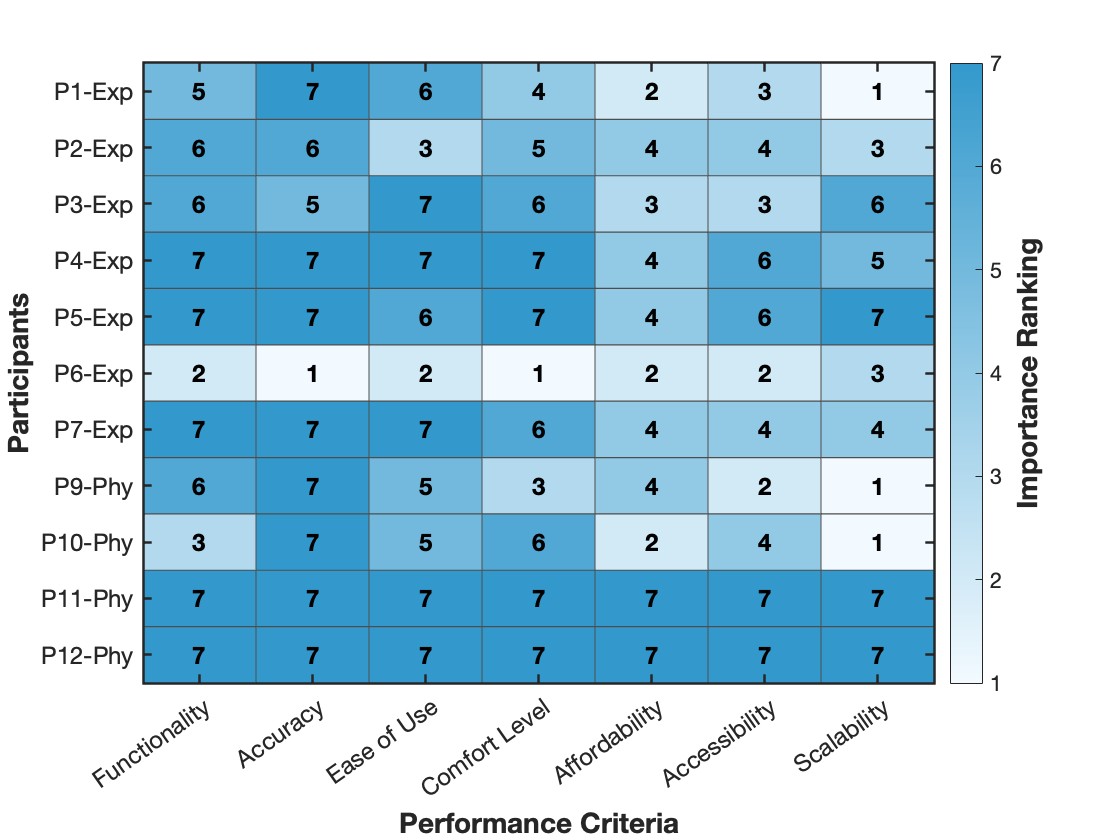}
    \caption{Importance rankings assigned by all participants to seven performance criteria for CGA sensing approaches (\textit{Darker shades indicate higher perceived importance. 
    Numerical scale of importance: 1 = no importance, 2 = slightly important, 
    3 = somewhat important, 4 = moderately important, 5 = important, 
    6 = very important, and 7 = extremely important}).}
    \label{fig:criteria_rank}
\end{figure}

\subsection{\textbf{RQ1: Clinicians' Perceptions of Computerized Gait Analysis Sensing Methods}}
This section explores clinicians’ perceptions of CGA sensing approaches. We report their ranking of sensor performance criteria and gait parameters, followed by medical experts' analysis of CGA sensing approaches. 

\subsubsection{\textbf{Clinicians Prioritize Accuracy, Usability, and Patient Comfort}}
Clinician participants ranked the importance of seven performance criteria on a scale from low to high importance. Overall, accuracy ($M$ = 6.18, $SD$ = 1.83) received the highest importance ratings, followed by functionality ($M$ = 5.72, $SD$ = 1.74), ease of use ($M$ = 5.64, $SD$ = 1.75), and comfort level ($M$ = 5.36, $SD$ = 1.96). These criteria were consistently rated highly across most participants. In contrast, affordability ($M$ = 3.91, $SD$ = 1.76), accessibility ($M$ = 4.36, $SD$ = 1.86), and scalability ($M$ = 4.09, $SD$ = 2.47) showed greater variability and generally lower importance ratings. This pattern suggests that clinicians prioritize technical performance and human factors over implementation-related considerations such as affordability, accessibility, and scalability when evaluating sensing approaches (Figure \ref{fig:criteria_rank}).

\begin{figure}
    \centering
    \includegraphics[width=1.10\linewidth]{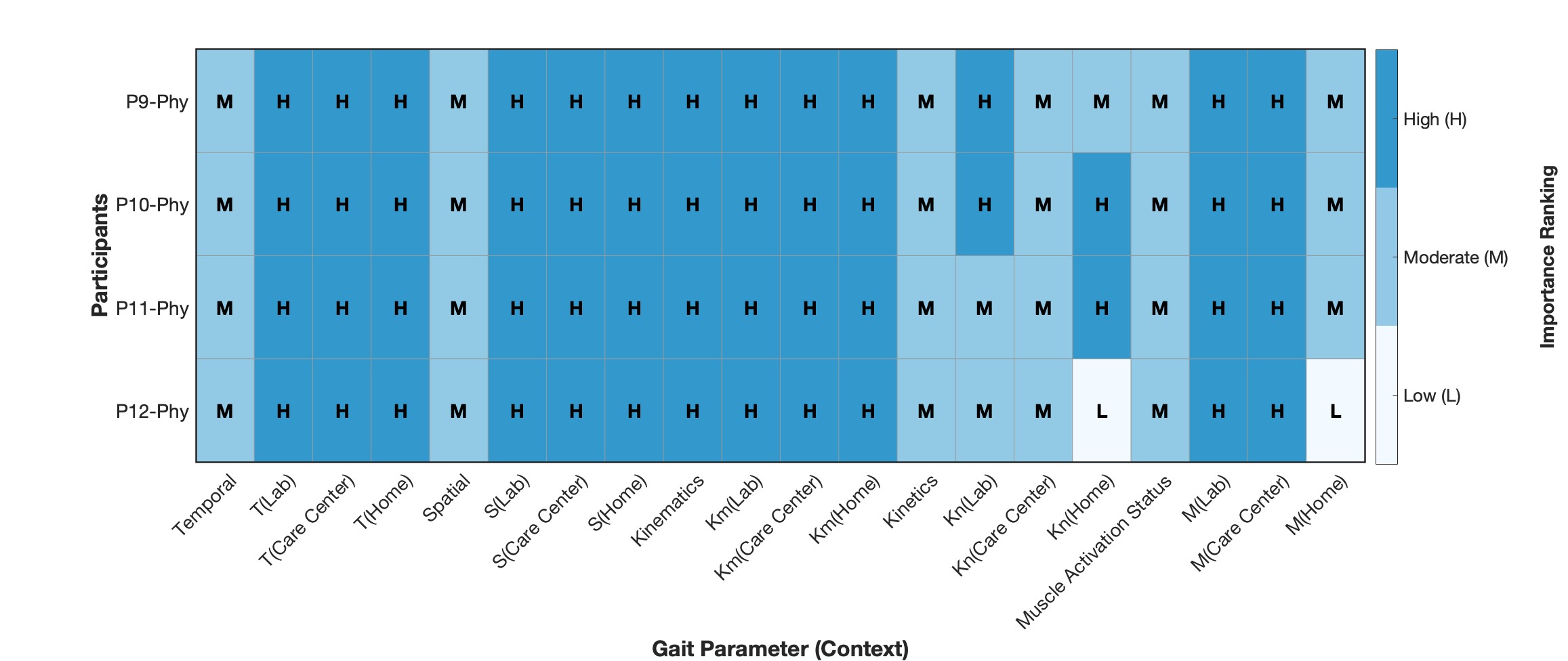}
    \caption{Importance rankings assigned by physician participants to five gait parameters across three contexts: Gait Lab, Local Care Center, and Patient's Home (\textit{Darker shades indicate higher perceived importance Numerical scale of importance: L = low importance, M = moderately important, H = high importance}).}
    \label{fig:gait_rank}
\end{figure}

Physicians were also asked to rank the importance of five gait parameters across different contexts (Figure \ref{fig:gait_rank}). Overall, kinematics was consistently rated as the most important parameter ($M$ = 3), receiving predominantly high importance ratings. However, the relative importance of the other parameters, temporal, spatial, kinetic, and muscle activation, varied depending on the analysis context (e.g., lab, care center, or home). This variation highlights that while kinematics is broadly prioritized, clinicians adapt their emphasis on other gait parameters based on the specific evaluation care setting. 

\subsubsection{\textbf{Sensing Preferences Vary Across Care Settings}}
Medical experts were asked to assess the 14 CGA sensing approaches (see Figure \ref{fig:approaches}) in the seven performance criteria. These participants, who are typically more technically situated, generally perceived most approaches as appropriate for capturing temporal and spatial parameters (64.3\% yes, 23.9\% not familiar, 11.7\% no), but fewer were considered capable of accurately measuring kinematics, kinetics, or muscle activation status. The findings also revealed limited familiarity with the radio frequency–based approach among these participants, which is an emerging wireless technology for human motion monitoring (Table \ref{tab:most_frequent_response})~\cite{wirelessgait2025}. 

\begin{figure}[H] \centering \includegraphics[width=0.8\linewidth]{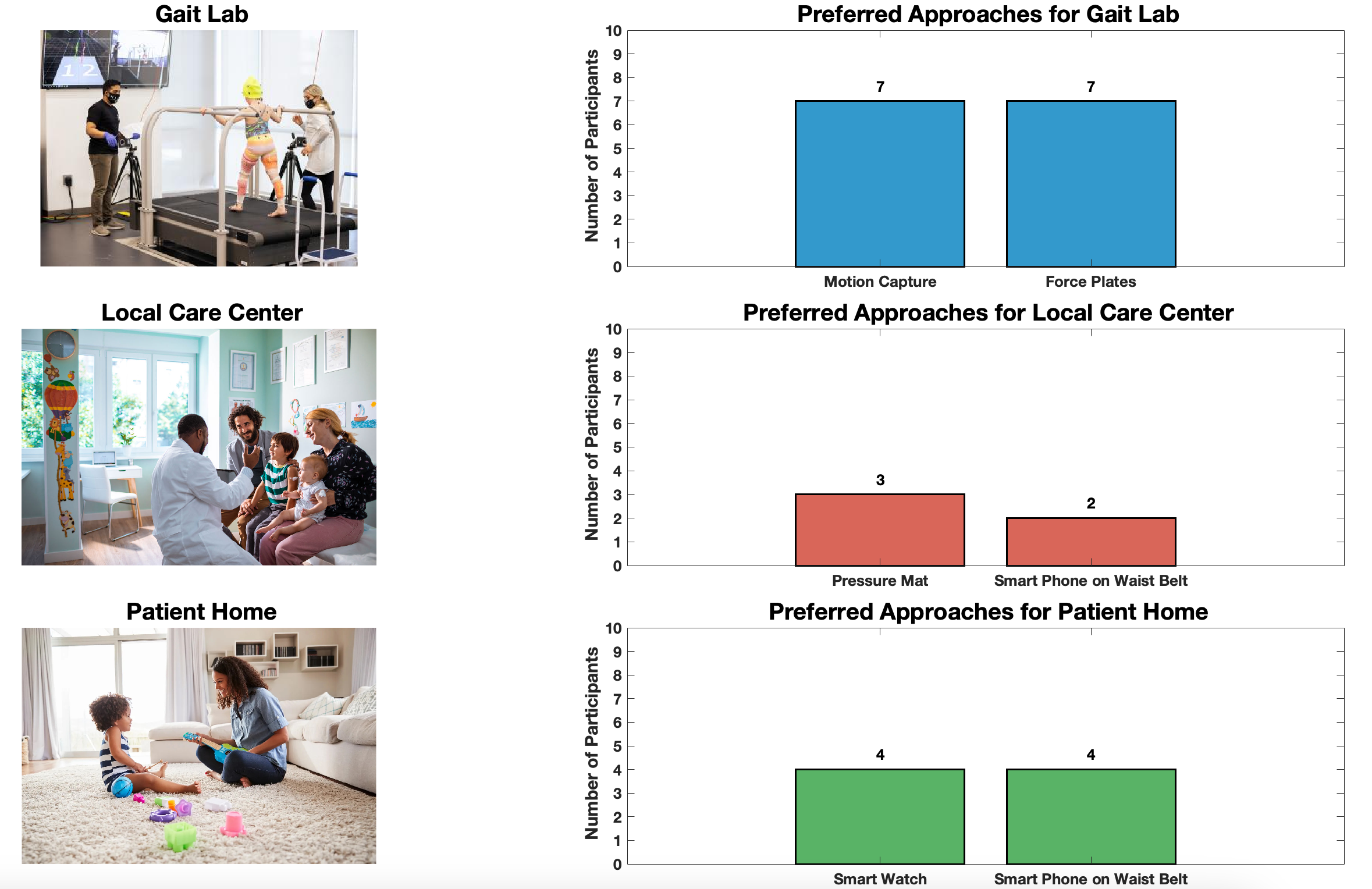} \caption{Images of three contexts (gait lab, local care center, and patient home) that were shown to medical expert participants (P1--P7). Participants indicated which gait analysis technology they would `definitely use'' in each context. The two sensing approaches with the highest number of `definitely use'' responses for each context are displayed in the corresponding bar charts on the right.} \label{fig:sensor_settings} \end{figure}

In addition, medical experts identified specific contexts in which each approach would be most appropriately used (Figure \ref{fig:sensor_settings}). Motion capture systems and force plates were predominantly associated with gait laboratory environments, reflecting their technical complexity and infrastructure requirements. In contrast, pressure mats and smartphones worn on a waist belt were more frequently linked to local care centers, where moderate levels of instrumentation are feasible. For home-based monitoring, wearable devices such as smartwatches and smartphones on a waist belt were most commonly identified as suitable options. These findings suggest that clinicians align sensing selection with the practical constraints and demands of each context, favoring highly instrumented systems in controlled environments and more portable, user-friendly solutions in decentralized care contexts. Notably, single video camera approaches appeared across multiple settings, suggesting that vision-based systems occupy a transitional role: familiar and accessible enough for varied contexts, yet not dominant in any one. The relatively low absolute counts for top-ranked approaches across all three contexts also reflect variation in clinician preferences, indicating that no single sensing approach has a clear consensus and that context-sensitive flexibility in system design may be more important than optimizing for a single dominant modality.

\subsection{\textbf{RQ1: How Clinicians Adapt, Interpret, and Negotiate Computerized Gait Analysis}}
We present participants' experiences adapting, interpreting, and negotiating CGA sensing approaches with child patients: 1) CGA expands and standardizes clinical evidence; (2) standardized data still requires clinical interpretation; and (3) pediatric care reveals mismatches between sensing systems and children’s bodies. 

\begin{table*}[t]
\centering
\caption{Most frequent response across medical expert participants ($n$=7)}
\label{tab:most_frequent_response}

\small
\setlength{\tabcolsep}{8pt}
\renewcommand{\arraystretch}{1.1}

\begin{tabular}{lccccc}
\toprule
\textbf{Technology} & \textbf{Temporal} & \textbf{Spatial} &
\textbf{Kinematics} & \textbf{Kinetics} &
\textbf{Muscle Activation} \\
\midrule
Motion Capture          & \checkmark & \checkmark & \checkmark & \checkmark & $\times$ \\
Force Plates            & \checkmark & $\times$ & $\times$ & \checkmark & $\times$ \\
EMG                     & $\times$ & $\times$ & $\times$ & $\times$ & \checkmark \\
Pressure Insoles        & \checkmark & \checkmark & $\times$ & -- & $\times$ \\
Pressure Mat            & \checkmark & \checkmark & $\times$ & $\times$ & $\times$ \\
Single Video Camera     & \checkmark & \checkmark & \checkmark & $\times$ & $\times$ \\
Multiple Video Camera   & \checkmark & \checkmark & \checkmark & $\times$ & $\times$ \\
Floor Vibration Sensors & \checkmark & \checkmark & -- & -- & -- \\
Radio Frequency         & -- & -- & -- & -- & -- \\
IMU                     & \checkmark & \checkmark & \checkmark & -- & $\times$ \\
Smart Watch (with IMU)  & \checkmark & \checkmark & $\times$ & $\times$ & $\times$ \\
Smart Phone (with IMU)  & \checkmark & \checkmark & $\times$ & $\times$ & $\times$ \\
Smart Glasses (with IMU)& \checkmark & \checkmark & $\times$ & $\times$ & $\times$ \\
Ear Pods (with IMU)     & -- & -- & $\times$ & -- & $\times$ \\
\bottomrule
\end{tabular}

\vspace{0.5em}

\begin{minipage}{\textwidth}
\footnotesize
\textbf{Abbreviations:}
Temp = Temporal Parameters;
Spat = Spatial Parameters;
Kin = Joint Angles (Kinematics);
GRF = Ground Reaction Forces (Kinetics);
Muscle Act. = Muscle Activation Status.

\textbf{Legend:}
\checkmark\ = Yes;
$\times$ = No;
-- = Not familiar.
\end{minipage}

\end{table*}

\subsubsection{\textbf{CGA Expands and Standardizes Clinical Evidence}}
Medical experts and the designer described how CGA has reshaped their interaction with the gait analysis process. A central shift they highlighted is the ability to ``\textit{extract a lot more information than a lot of physicians were traditionally used to having}'' (P13-Des). Similarly, P2-Exp noted, ``\textit{A lot has changed in the last five years... now we are able to get pretty good [information] in the sagittal plane}.'' Reflecting on this shift, P1-Exp emphasized how access to more comprehensive data supports clinical decision-making: ``\textit{Oftentimes the decisions that are coming out of these gait analysis appointments are really life-changing decisions, so to have a little bit more data, a little bit more information to make those decisions is really important}.''

Beyond enabling more detailed analysis, participants also described these objective measures as critical for communicating clinical value across stakeholders. P3-Exp explained that ``\textit{quantifiable measures}'' help communicate progress not only to physicians and families but also to payers, thereby strengthening the case for insurance coverage of interventions. In a similar vein, P5-Exp noted that computerized methods support structured before-and-after comparisons and provide a more scalable basis for evaluation, even as visual interpretations remain important for clinical communication. P2-Exp further highlighted the shift introduced by these systems, noting that computerized methods produce ``\textit{objective}'' data compared to traditional visual analysis, which can vary across clinicians---thereby standardizing what was previously subjective clinical judgment.

Physician participants similarly expressed receptiveness to these capabilities, particularly the standardization and expanded analytic scope afforded by computational approaches. P8-Phy emphasized that ``\textit{having a standardized way of looking at a patient is really important},'' underscoring the value of consistency in clinical interpretation. P9 also stated, ``\textit{I hope that there is slightly more [objectivity] rather than traditional methods of recording, which I think sometimes can be less accurate}.'' P8-Phy reflected on how technological advances have expanded what can be measured and analyzed: ``\textit{Those graphs are incredibly helpful in sort of understanding at the level of what's happening... because you can measure angles while they're walking, which is really hard to do if you don't have some sort of video-based analysis. And so you don't know if they're using their entire range}.'' 

P11-Phy also stated:

\begin{quote}
``\textit{I remember in the early days, the sensors were hardwired directly to the machine, so you had to [analyze] on a treadmill. One of the researchers [now] has his own separate lab with microscopic sensors that can look at individual muscles. [The downsizing of technology] has made for a better ability to assess muscle function, a better ability to figure out what the true kinematics are... [and] how much a knee is moving. Now, you can see the movements in all three planes, so things are much more accurate than they used to be}.''
\end{quote}

Similarly, P10-Phy described how computerized systems extend clinicians' perceptual and analytical capacity:

\begin{quote}
``\textit{I think that the clinical evaluation is quite limited based on just space and ability to see an individual. A three-dimensional image and three-dimensional motion analysis really give a better opportunity to see all the different planes. I think also just being able to have objective numbers to look at can be really helpful too in terms of surgical decision-making and how much rotation to make}.''
\end{quote}

These accounts illustrate how CGA mediates clinical work by transforming movement into analyzable data, enabling richer and more standardized forms of evidence that can strengthen---or challenge---clinical decision-making around interventions.

Additionally, unlike visual observation, which depends on real-time human judgment, CGA can capture data first and analyze it later, enabling new modes of review and comparison. P13-Des stated, ``\textit{You can have someone record the analysis and then watch it later. So an expert can have a look at it later, and they can also play it faster. This sounds like a simple thing, but that actually saves clinicians time}.'' Complementarily, P8-Phy stated that ``\textit{having a computer that does [the work] for you is great, or at least takes a significant portion of that burden off a person}.'' 

\subsubsection{\textbf{Standardized Data Still Requires Clinical Interpretation}}
However, a few physicians noted that the standardization of computerized methods introduces new dependencies and does not fully eliminate subjectivity. P9-Phy observed that uncertainty can persist even within digital systems, stating: ``\textit{I think even with digital, there's still a lot of human interpretation, so it's still very visual in some ways. I don't think some of the measurements are completely automated}.'' Extending this concern, P8-Phy reflected on the limitations of current systems in supporting complex clinical decision-making, particularly in cases where clear recommendations are difficult to derive or standardization is challenging: ``\textit{I went through seven gait analyses yesterday. And there [were] a couple patients where I could recommend something, but I don't really know if that's going to make that specific patient better because they have dyskinetic features\footnote{People with dyskinetic cerebral palsy experience involuntary, fluctuating movements that occur outside of their control \cite{CPARF_DyskineticCerebralPalsy}.}... we don’t have those answers}.'' This uncertainty can also stem from the sensors themselves, as P13-Des noted that ``\textit{each modality has its issues},'' indicating that the limitations of individual sensing modalities can shape what clinicians are able to interpret from the resulting data. These perspectives show that CGA data do not produce clinical meaning on their own. Clinicians must interpret standardized measurements in relation to the individual patient and technical limitations. When clinically relevant phenomena, such as dyskinetic movement patterns, are not captured, uncertainty is shifted back onto the clinician rather than being resolved or mitigated by the technology. Thus, while CGA can standardize how gait is measured, translating those measurements into clinical meaning remains a situated process.

\subsubsection{\textbf{Pediatric Care Reveals Mismatches Between Sensing Systems and Children's Bodies}}
While CGA provides objective data, its sensing systems introduce challenges when interacting directly with child patients. The findings show how pediatric CGA is shaped by both the requirements of sensors and children's bodies, sensory needs, and participation. Clinicians and care teams often adapt sensing procedures to make these systems work with children, highlighting ways in which current sensors do not fully account for the needs of pediatric care.

\paragraph{\textbf{Challenges of Sensor Placement on Children’s Bodies}}
Pediatric anatomy introduces calibration and configuration burdens, revealing mismatches between system design assumptions and actual use contexts. As medical experts primarily navigate conducting the gait analysis, they describe how these mismatches materialize in practice. For example, P1-Exp reflected on anatomical differences in children and their impact on placement and measurement accuracy: ``\textit{Sensor placement sometimes is difficult on a really small body or individuals who have bony abnormalities\footnote{A bony abnormality is a deviation from normal bone structure \cite{michaels2012ear}.}}. She further stated, ``\textit{There are rotational deformities in bones, or different, harder-to-find bony landmarks [which] can be tough. Maybe [for] patients who are a little bit more overweight, it's harder to find some of those bony landmarks. So I think that can impact the accuracy}.'' P12-Phy also supported this perspective:

\begin{quote}
``\textit{Kids who have a lot of abnormality in their movement or their bone structure can knock off these markers. Current, state-of-the-art digital motion capture requires light reflective markers placed on bony landmarks. And, you know, there's noise and potential error in the placement of those markers on bony landmarks. Obviously, the amount of sub-tissue underneath them makes a difference}.''
\end{quote}

Beyond anatomical variation, participants also emphasized how developmental and sensory differences shape children’s interaction with sensing equipment. P4-Exp described how some children with cognitive delays may actively remove sensors, while others respond strongly due to sensory sensitivities, as P1 stated: ``\textit{We get a lot of patients who have sensory processing disorders, so having things put on their bodies is really uncomfortable for them. [As a result], sometimes we have sensors thrown across the room}.'' 

These challenges also shape methodological decision-making. P5-Exp explained that sensor choice depends heavily on patient condition, particularly for children with neurological conditions who may be more sensitive to sensors placed on their bodies: ``\textit{For [the] pediatric population, [it] depends on what condition they have. If it’s more like orthopedic or sports-relevant injuries, there’s probably not too much of a deal for the kids. But if it’s more like neurological conditions, everything on their body can be really sensitive... You want to be as minimal as possible and without interfering with the walking patterns if you do a gait analysis}.'' At the same time, she highlighted an inherent constraint: certain clinically necessary measurements, such as muscle activation, require specific sensors (e.g., EMG), limiting flexibility in system design.

These perspectives reveal a tension between collecting the data needed for clinical assessment and minimizing the sensing burden on children. While some measurements require multiple or contact-based sensors, children may not always tolerate these setups. Clinicians therefore described adapting the process to help children become more comfortable with sensors. For example, P4-Exp described preparing children beforehand using stickers or Band-Aids: ``\textit{I have them stick Band-Aids or stickers onto different parts of the body beforehand just to get used to that feeling of having that on and just ripping it off as well}.'' These strategies show that making sensors work with children can require additional preparation and coordination beyond the technology itself. 

\paragraph{\textbf{Keeping Sensors in Place During Pediatric CGA}}
Children's movement can also make it difficult to keep sensors in place throughout an assessment. P5-Exp explained that pediatric CGA may involve activities beyond walking, such as jumping and landing, which can cause children to sweat and markers to become dislodged: ``\textit{When they start to move more, they start to sweat, and then the marker can fall off really easily. We try tape, like FebriTape, and use those [adhesive] sprays so they can stay better}.''

Similarly, P6-Exp observed that markers ``\textit{may not end up in the exact same spot as they were when we first started},'' introducing variability into the data. Maintaining consistent marker placement can therefore be difficult throughout pediatric CGA, particularly as children move and sensors become displaced. P3-Exp further reinforced this, noting that variability in adhesion across sessions can degrade data quality. When markers become loose or fall off, clinicians and technicians may need to reapply them, recalibrate the system, and repeat parts of the data collection process.

\paragraph{\textbf{Adapting CGA to Children’s Participation}}
Medical expert participants also described how children's participation can shape the CGA process. P3-Exp noted, ``\textit{Working with kids and wearable sensors or getting them to walk straight on a gait mat can be unpredictable at best},'' explaining that even calibration can be difficult when children have difficulty remaining still: ``\textit{Just getting the kids, especially younger kids, to stand still to calibrate the system that we were using at the time}.'' P1-Exp further described how patient endurance can affect the assessment process: ``\textit{Some of those markers do have to be put on in weight-bearing positions… If our patient requires rest breaks every minute or so, it just takes a lot more time... and the longer it takes and the more energy they expend putting the markers on, the less energy they have for the analysis piece}.'' These experiences show how data collection procedures may need to be adjusted around children's ability to remain still, follow measurement procedures, and sustain activity throughout the assessment. 

\paragraph{\textbf{Adapting Pediatric CGA Through Team-Based Care}}
Physicians also described how pediatric CGA often requires support from multiple members of the care team to help children navigate the gait analysis experience. This can involve child life specialists and other support staff who use toys, games, and other strategies to make the assessment more comfortable and engaging for children. For example, P8-Phy described:

\begin{quote}
``\textit{I will say if you had a child life specialist with you, that would probably be way better. We have flashy toys and [such] for the younger kids. Making it a game has [also] been helpful. You can have them take a picture with a flash on, so they look like a robot and see themselves like that}.''
\end{quote}

P10-Phy also highlighted interaction strategies, ``\textit{Playing with them [helps]. In the clinic, we have a little basketball hoop with a little soft foam ball that they really gravitate toward}.'' Similarly, P11-Phy stated, ``\textit{I always kept toys in my pockets for the little kids}.'' These strategies adapt the assessment experience around the child, using play and additional clinical support to help children become comfortable with unfamiliar sensors and procedures. These practices also suggest opportunities for CGA to better support child-friendly strategies, such as play and interactive feedback, as part of the process. 

The role of the care team can also extend beyond making the analysis more comfortable and engaging. P8-Phy explained that some children may require more support throughout the evaluation process, noting that for children with more significant developmental delay or autism, support staff often take on much of the ``\textit{legwork}'' in guiding them through the analysis. Other members of the care team also contribute knowledge that helps clinicians understand the child beyond what is captured during the analysis. P10-Phy explained:

\begin{quote}
``\textit{Having a team of physical therapists and occupational therapists working with those kids, they become sort of the better source of truth for what the kids are actually like}.''
\end{quote}

These perspectives show that pediatric CGA can involve different forms of support across the care team, from helping children through the assessment to providing contextual knowledge about their movement and behavior outside the gait lab. This team-based support helps clinicians adapt the analysis process to children's individual needs and provides context for understanding the data collected during the assessment.

\subsection{\textbf{RQ2: Design Requirements for Pediatric Computerized Gait Analysis}}
All participants articulated visions for design improvements that would better serve children while maintaining clinical utility and increasing accessibility. In discussing future directions, P1-Exp emphasized the importance of advancing CGA toward more personalized, accessible, and human-centered models of care: ``\textit{We get many parents who didn’t even know this kind of resource existed. If we can help them recognize that our services are available—-that we can provide support and ideally make their appointments as comfortable and low-stress as possible-—that would be my ideal outcome. I want both patients and parents to feel heard, seen, and assured that this is an analysis designed for them, where they remain our top priority}.''

\subsubsection{\textbf{Reducing Sensing Burden While Preserving Clinical Value}}
Across participants, there was a shared interest in moving from contact-based sensing toward more unobtrusive approaches. Medical experts saw markerless and reduced-contact sensing as a way to reduce the physical and sensory burden on children while also addressing challenges associated with placing sensors on their bodies. Participants therefore emphasized minimizing the number of markers or sensors where possible. As P1-Exp noted, ``\textit{if it could be something that has [as] minimum markers as possible to put on their body. That would be wonderful},'' particularly when working with children who may be anxious or sensitive to sensors.

P4-Exp similarly described markerless motion capture as ``\textit{the gait analysis of the future},'' highlighting its potential to reduce ``\textit{any type of medical trauma inflicted on these children}.'' He also explained that reducing physical contact could improve data quality by allowing children to walk more naturally:

\begin{quote}
``\textit{The less amount of touch that you have on these patients, I think the better data you're probably going to get... [because] a lot of these patients, if they're not used to having all these things on them, they're not going to walk as they typically do for your analysis}.''
\end{quote}

Reducing contact may therefore not only make gait analysis more comfortable for children, but also help capture movement that better reflects how they typically walk. However, participants recognized that reducing contact is not possible for every type of gait measurement. P13-Des explained that certain physiological signals still require direct sensing, noting: ``\textit{You have to measure the nerve. I don't see a way you would do that without an EMG-type sensor}.'' Thus, participants wanted systems that minimize contact where possible without removing sensors needed to capture clinically important measurements.

\subsubsection{\textbf{Fitting Gait Analysis into Clinical Workflows and Environments}}
Participants emphasized that future CGA should also be easy for clinicians to use. As P3-Exp explained, simplicity is particularly important when clinicians have limited time with patients: ``\textit{Simplicity is king. The simpler, the better. As a care provider, we have a very limited amount of time with the patient; we need something quick, easy, reliable, and consistent. We need it to be a point-and-click solution. I need to be able to just press a button and have it give me the information that I need}.'' P11-Phy similarly wanted future systems to be ``\textit{extremely easy to operate},'' while P10-Phy preferred a single technology that could provide ``\textit{all that information in a sort of neat little package}'' rather than requiring multiple sensing systems. These perspectives suggest that reducing setup and operation demands is important for fitting CGA into clinical workflows.

Participants also described how the clinical environment can affect children's experience during gait analysis. P1-Exp explained that children may enter the gait lab already stressed because ``\textit{it looks very sterile... that stops them at the door immediately}.'' Even routine preparation can create negative associations, as ``\textit{the smell reminds them of shots, so they get stressed already from that}.'' P1-Exp suggested creating a more ``\textit{warm and friendly and less sterile}'' environment, contrasting the gait lab with spaces designed specifically for children: ``\textit{If you walk into the children's hospital, there are pictures and animals and trains everywhere. It's very different than the adult hospital for a reason}.''

P8-Phy similarly emphasized the importance of the physical space, noting that assessments can benefit from ``\textit{a quiet, enclosed space}'' with fewer distractions, although such spaces may not always be available because of financial and logistical constraints. These experiences suggest that designing pediatric CGA requires attention not only to the sensors but also to the clinical environment in which measurement takes place.

\subsubsection{\textbf{Supporting Gait Measurement Across Patients and Settings}}
Participants also wanted sensors that could accommodate different patients, activities, and settings. P8-Phy, for example, emphasized the need for sensors that can work with assistive devices such as braces without compromising measurement. Participants also saw opportunities to move gait measurement beyond a single assessment in the gait lab. P9-Phy suggested that more accessible sensors, such as a phone-based application, could allow patients to collect movement data over time. They explained how longitudinal measurement could support return-to-play decisions:

\begin{quote}
``\textit{Ideally, it would allow me to determine if someone’s actually truly ready to return at full capacity. If we could do some type of analysis during practice, but not on a single day, but every single practice, we can see consistently, is this athlete or this patient ready to perform at the levels that we intend them to? And perhaps we can use it to capture baseline pre-injury}.''
\end{quote}

Participants also considered ambient sensing as a way to make data collection less noticeable. P12-Phy suggested that ``\textit{the less noticeable, the better},'' proposing micro-vibration sensors that could be placed around a room without being noticeable to patients. However, less noticeable sensing would still need to capture clinically important measurements while addressing concerns such as privacy. P12-Phy explained:

\begin{quote}
``\textit{I would design a sensor that could pick up limb motion and joint kinematics but wouldn't invade privacy. I don't know what the technology would be, whether it would be radar or some video kind of blurred pixelation, not just of the face, because [it is] very hard to actually de-identify somebody and keep the blur on their face. Those automatic face blurring technologies definitely do not keep up with somebody walking at this point, though they claim to}.''
\end{quote}

Participants envisioned CGA that could collect clinically useful gait information across a wider range of patients, activities, and settings. This includes accommodating assistive devices, supporting repeated measurements outside a single clinical visit, and exploring less noticeable forms of sensing. At the same time, these approaches still need to provide the measurements clinicians need while protecting patient privacy.

\subsubsection{\textbf{Extending Gait Analysis to Children's Everyday Environments}}
Participants envisioned future CGA extending beyond specialized gait laboratories into the environments where children naturally move. They described this as an opportunity to capture movement that may not be visible during a single clinical assessment and to understand how children move across activities and over time.

\paragraph{\textbf{Capturing Movement in Everyday Settings.}}
Medical expert participants emphasized the value of assessing movement in environments that reflect children's everyday activities. P2-Exp described the importance of recreating relevant environments for athletes:

\begin{quote}
``\textit{Having a space that recreates their sporting environment would be the best or most ideal... being able to meet the patients in their environments would be better, so portability of the systems to make that happen}.''
\end{quote}

P5-Exp similarly highlighted opportunities to observe children across settings such as ``\textit{a school setting... an outdoor setting, [or] a playground}.'' These settings could provide information about how children move during activities that may be difficult to reproduce within a gait lab. P3-Exp contrasted this type of measurement with the snapshot provided by a clinical assessment:

\begin{quote}
``\textit{Data collected in [the] lab is going to be so much different than data collected out of [the] lab. A movement library of them over time outside of the office would be really interesting... versus the way they’re moving in the office, which is kind of just a snapshot in time}.''
\end{quote}

Participants further highlighted the value of extending CGA into everyday environments, as children's movement in the gait lab may not always reflect how they typically move. P10-Phy described how being observed during a clinical assessment can change a child's walking:

\begin{quote}
``\textit{If they know they're being watched, they walk differently. Every now and then, we will see someone in our motion lab, and they will walk beautifully. [However], that is not how that kid walks. If that kid walked that way, I would never have sent them to the motion lab. So I think that, you know, there's something artificial that's created when you send the person into a medical facility to walk. I think something like at home and watching them walk would be more helpful}.''
\end{quote}

These findings suggest that extending gait analysis beyond the laboratory could help clinicians capture movement in settings and activities that more closely reflect how children move in their everyday lives.

\paragraph{\textbf{Supporting Longitudinal Measurement Outside the Clinic}}
Participants also saw value in collecting gait and movement data over time rather than relying on measurements from isolated clinical visits. P12-Phy distinguished between the information provided by gait laboratories and the potential role of home monitoring:

\begin{quote}
``\textit{Physical and functional analyses [are] well-suited for [the clinic and gait lab]. I think home monitoring has a different goal, and that would be risk analysis for trips and falls, to make sure that patients are safe in their homes, [and activity monitoring]. Then you can see if there's a change in the gait pattern that's sudden, and that would be important information to know to avoid trips and falls, but also to identify changes in health status and neurologic status}.''
\end{quote}

Instead of replacing laboratory assessment, participants described home and everyday sensing as providing different information about changes in movement, safety, and recovery over time. P9-Phy explained, ``\textit{If we were able to track someone's recovery, not just during a single session, but [daily] through their practice, then that becomes a much more powerful tool}.'' P10-Phy similarly wanted longitudinal measures of outcomes that are directly relevant to children and their families, such as how often a child trips and whether mobility changes following treatment:

\begin{quote}
``\textit{I would like to know [what the actual frequency of tripping is]? It's a very valid sort of analysis [because] it's something that could cause harm [and] something that parents are worried about... I think there's probably some opportunity to better understand pre-op and post-op. Are we making these kids better? Do they walk faster}?''
\end{quote}

These perspectives show how extending CGA beyond individual clinical visits could support different forms of longitudinal assessment, including monitoring recovery, changes in gait, fall risk, and outcomes that matter for child development.

\paragraph{\textbf{Designing Sensing for Use Outside the Gait Lab}}
Moving gait analysis into everyday environments also changes the requirements of the sensors themselves. P13-Des described the potential for measuring gait and balance in free-living settings while noting that sensor selection depends on where and how the technology will be used:

\begin{quote}
``\textit{We can move outside the clinic and start measuring gait and balance and things in the free-living sort of context. There are things like privacy and other considerations that play into picking which sensors you want to use and then what kind of scenarios you can use which sensor}.''
\end{quote}

P13-Des described smartphones as one possible way to make home-based measurement more feasible without requiring additional sensing infrastructure:

\begin{quote}
``\textit{We and other people are developing a smartphone camera to have people perform specific standardized tests. A way of doing the computer vision part at home without additional hardware or privacy concerns, such as putting cameras in their homes. It's largely driven by [the fact that] if you don't use consumer devices, then it becomes much more difficult to actually implement and scale things}.''
\end{quote}

At the same time, P13-Des noted that moving sensing outside the gait lab introduces practical limitations. Sensors can remain fixed and configured in dedicated clinical spaces, whereas deploying them across homes and other environments can require additional hardware and setup. Although smartphone-based approaches may make data collection more accessible, some forms of computer vision still benefit from multiple cameras to reduce occlusion and improve measurement quality. Overall, participants envisioned extending pediatric CGA beyond individual assessments in specialized gait laboratories. Schools, playgrounds, homes, and sports settings could provide information about how children move during everyday activities, while repeated measurements could help clinicians understand changes over time. However, moving sensing into these environments also introduces requirements around portability, privacy, setup, and measurement quality. 

\subsubsection{\textbf{Decreasing Clinical Work Through Automation}}
Participants described automation as a way to reduce the time and manual work involved in processing gait data. P8-Phy described the biomechanist role as particularly time- and resource-intensive: ``\textit{The biomechanist role [is] a lot of work. Going through and cleaning up the data and all that... Those people are very expensive, and the amount of time that's required is incredibly high}.'' Similarly, P9-Phy explained that automation could make the CGA process more efficient: ``\textit{If it were automated, it would actually speed up the process quite a bit. I think the team, as far as I know, spends a lot of time processing the data, which, candidly, won't be cost-effective for the hospital system}.'' Participants also saw opportunities for automation to support more frequent monitoring. P12-Phy expressed interest in an ``\textit{automated report that the clinicians could use to keep monitoring multiple times a year},'' which could make it easier to track patients over time without requiring the same amount of manual processing for each assessment. At the same time, one participant recognized that greater automation could change existing clinical roles. P2-Exp, a biomechanist, explained:

\begin{quote}
``\textit{I'm going to say my role is going to be diminished, especially with the advent of the video-based systems, because there [are] less requirements for technical expertise to operate those [and] process the data, and some of that's becoming more automated}.''
\end{quote}

These perspectives show that automation could reduce the work required to operate CGA and process gait data while also changing the role of the clinical staff who currently perform this work.

\section{Discussion}
We conducted a mixed-methods study with pediatric clinicians and a system designer to understand how they perceive, interpret, adapt, and negotiate CGA when working with child patients, as well as how they envision future CGA systems. In this section, we discuss our findings and the design considerations that emerge from them: 1) aligning CGA with children's bodies and needs, 2) the care setting as part of CGA, and 3) automating the work around pediatric CGA. Lastly, we present design guidelines for pediatric CGA and design implications for designing healthcare technologies for pediatric populations.

\subsection{Aligning CGA with Children's Bodies and Needs}
Our findings show that designing CGA around children's needs is important for both the child's experience and the gait data being collected. Participants described how children's bodies, sensory needs, fatigue, movement, and ability to participate can shape the measurement process. Sensors can be difficult to place on smaller bodies, markers can move or fall off during activity, and contact-based sensors can affect children's comfort and how they move. Children may also become tired during setup or have difficulty remaining still for calibration. Our participants noted that the interaction between the child and the sensing technology can impact whether data can be collected and whether the resulting measurements reflect the child's typical movement.

Prior work has shown that clinical assessment with children often needs to account for their developmental abilities and participation~\cite{srinath2019clinical}. Pediatric assessment guidelines describe the importance of interacting with children at their developmental level and using approaches such as play to support assessment~\cite{srinath2019clinical}. Research on wearable sensing with children has also identified challenges related to comfort, compliance, and individual needs~\cite{collier2025scoping}. Our findings build on this work by showing how these challenges impact CGA. For CGA, a child's ability to participate is closely connected to the data being produced. A marker that moves, difficulty remaining still during calibration, or discomfort that changes how a child walks can affect the resulting gait measurements.

Other areas of pediatric healthcare have also adapted technologies and assessment procedures to children's needs. Pediatric audiology has developed protocols that account for differences in children's anatomy, attention, and developmental abilities~\cite{bagatto2010protocol}. Pediatric cardiac monitoring similarly considers smaller electrode placement and movement artifacts~\cite{bonafide2013development}. Our findings identify similar issues in pediatric CGA and show how clinicians currently manage them during gait assessment. Clinicians described adjusting marker placement, reapplying displaced markers, repeating calibration or data collection, minimizing sensors when possible, preparing children for contact with sensors, and using play and additional clinical support during the assessment. This connects with HCI research showing how clinicians adapt technologies around patients, workflows, clinical judgment, and local conditions~\cite{beede2020human,sivaraman2023ignore,ng2019provider}. In pediatric CGA, this adaptation also occurs during data collection, where clinicians have to make systems work with children's bodies and participation. When CGA systems do not account for these needs, more of this work falls to clinicians and care teams.

Our findings point to several ways pediatric CGA could better support this work. Calibration and sensing methods could better accommodate smaller and changing bodies, sensors could be designed to remain reliable as children move, and physical contact could be reduced where possible. Some clinically important measurements may still require direct sensing, such as EMG for measuring muscle activity. The design challenge is therefore to reduce sensing burden while preserving the clinical information needed for gait assessment. Accounting for children's bodies, sensory needs, and participation in the design of CGA could reduce the adaptations required during assessment and support measurements that better reflect how children move.

\subsection{The Care Setting as Part of CGA}
Our findings show that the care setting where gait assessment takes place can shape both children's experience and the movement being measured. Participants described how sterile clinical spaces can make children uncomfortable before the assessment begins, while more child-friendly environments can help children feel more comfortable. This finding connects with prior work on pediatric healthcare environments showing that features such as color, lighting, and child-centered spatial design can impact children's experiences and stress in clinical settings~\cite{dalke2006colour,litwin2023designing}. Our findings extend these considerations to CGA, where the environment may also affect the clinical data being collected. Because movement is what CGA measures, changes in how a child moves in response to the clinical environment can become part of the resulting gait data.

Participants also questioned whether movement captured in a gait lab always reflects how children typically move. They described children walking differently when they know they are being observed and noted that a single assessment provides a limited view of movement over time. Prior work in pediatric rehabilitation has emphasized the importance of children's participation and activity within natural environments~\cite{hall2021pediatric,anaby2012predictors}. Our findings show how CGA could support this by extending gait measurement into homes, schools, playgrounds, and sports settings. These environments could provide information about how children move during everyday activities and allow clinicians to examine changes across time instead of relying on a single clinical assessment. Participants saw this as particularly useful for understanding recovery, changes in gait, tripping, and other outcomes that may be difficult to observe during one visit to the gait lab.

Extending CGA across these settings also changes what is required from the sensing technology. Participants discussed portable and smartphone-based approaches as ways to make gait measurement more feasible outside specialized laboratories. They also saw potential in ambient and vision-based sensing for reducing the amount of equipment placed on the child's body. These approaches introduce other concerns, especially around privacy, setup, occlusion, and measurement quality. Participants were particularly concerned about using cameras in homes and whether privacy protections such as face blurring would remain reliable as people move through the environment. These concerns connect with broader HCI work on sensing outside controlled clinical settings, where technologies must work within the conditions and routines of everyday environments~\cite{sun2024multimodal,alharbi2018can}.

Our findings therefore point to designing pediatric CGA across different care environments. A gait lab can provide controlled measurement, but its physical and social environment should also be designed around children and the ways they experience the assessment. Portable systems can support measurement in local care settings, while home and community sensing can provide information about movement across activities and over time. Each setting introduces different requirements for setup, privacy, portability, and measurement quality. For pediatric CGA, where sensing takes place can shape what movement is captured and what clinicians are able to learn from it. The setting should therefore be considered when designing for sensing performance and patient needs.

\subsection{Automating the Work Around Pediatric CGA}
Our findings show that automation could reduce some of the time and work required to use CGA in clinical practice. Participants described substantial work involved in cleaning and processing gait data, generating reports, and preparing measurements for clinical use. Physicians saw automation as a way to reduce this work, make gait analysis more feasible within clinical workflows, and support more frequent monitoring over time. Prior HCI work has similarly examined how automation can reduce clinical workload while raising questions about how computational systems change existing roles and responsibilities~\cite{sivaraman2023ignore}. Our findings add to this work by showing that the work surrounding pediatric CGA extends beyond processing the data after it has been collected.

As our earlier findings show, clinicians and care teams also perform additional work to make gait sensing function with children. They adjust sensor placement, reapply markers, repeat calibration, prepare children for sensors, use play and other strategies to support participation, and coordinate with other members of the care team. HCI research has described how clinicians adapt technologies when they do not fit existing workflows, patient needs, or local conditions~\cite{beede2020human,sivaraman2023ignore,ng2019provider}. Research on invisible work and workarounds has also shown how work needed to make technologies function in practice can fall outside the tasks that systems are designed to support~\cite{stisen2016accounting}. Our findings show this in pediatric CGA: some of the work clinicians perform comes from mismatches between the requirements of sensing systems and the children and environments in which those systems are used.

This factor changes where automation could provide support. Automating data processing and reporting can reduce work at one stage of the CGA process, but it does not address the work required to make data collection possible. CGA systems could also support clinicians in identifying displaced markers, determining when recalibration is needed, recognizing trials that may need to be repeated, or reducing the setup required before an assessment. The design opportunities discussed in the previous sections---such as lower-contact sensing, calibration that better accommodates children's bodies, and sensing that works across different environments---can also reduce work by addressing the conditions that create it.

Automation also needs to account for the interpretation that happens after gait data are collected. Participants described cases where standardized measurements did not provide a clear clinical recommendation and where clinicians needed to interpret the data in relation to the individual child. This connects with prior HCI work showing that clinicians do not passively accept computational outputs; they interpret them using clinical judgment and knowledge of the patient and care context~\cite{sivaraman2023ignore,beede2020human,ng2019provider}. Prior work on gait analysis has also identified data complexity and interpretation as barriers to clinical use~\cite{obrien2021not,hebda2022clinicians}, while the presentation of gait data can affect how clinicians understand and use it~\cite{seals2022they}. Automation of processing and reporting therefore still needs to provide clinicians with enough information to understand the resulting measurements and determine how they apply to the child.

Our findings also show that increasing automation may change existing clinical roles. One biomechanist anticipated that greater use of video-based systems and automated processing could reduce the need for some of the technical expertise currently involved in CGA. This makes automation a question of both reducing work and understanding which forms of work are changing. For pediatric CGA, automation could be most useful when it reduces repetitive processing and the additional work created by current system requirements while continuing to support the clinical interpretation needed to understand what gait measurements mean for an individual child.

\subsection{Design Guidelines and Implications}
Participants revealed mismatches between current gait technologies and the needs of pediatric care. Table~\ref{tab:design_implications} summarizes these mismatches, the adaptations clinicians currently make, the resulting design guidelines, and design implications for future pediatric healthcare technologies.

\begin{table*}
\centering
\caption{Design guidelines from observed mismatches in pediatric CGA to broader implications for pediatric healthcare technology}
\label{tab:design_implications}

\renewcommand{\arraystretch}{1.15}

\begin{tabularx}{\linewidth}{
>{\RaggedRight\arraybackslash}p{0.22\linewidth}
>{\RaggedRight\arraybackslash}p{0.22\linewidth}
>{\RaggedRight\arraybackslash}p{0.24\linewidth}
>{\RaggedRight\arraybackslash}X}
\toprule

\textbf{Observed mismatch} &
\textbf{Current clinical adaptation} &
\textbf{Design guidelines for pediatric CGA} &
\textbf{Design implications for pediatric healthcare technology} \\

\midrule

Sensor placement and calibration can be difficult on children's smaller and variable bodies &
Clinicians adjust marker placement, recalibrate systems, and repeat measurements when markers move or become displaced &
Design sensing and calibration methods that accommodate smaller and changing bodies and remain reliable during movement &
Design pediatric technologies around variation in children's bodies rather than requiring clinicians to continually adapt standardized systems \\

\midrule

Contact-based sensing can be uncomfortable for children and may affect how they move during an assessment &
Clinicians minimize sensors where possible and prepare children beforehand using strategies such as stickers or Band-Aids &
Reduce physical contact through markerless or minimal-contact sensing while preserving clinically necessary measurements &
Treat children's comfort and sensory needs as requirements that can directly shape the quality and validity of clinical data \\

\midrule

Standardized data collection procedures do not always align with children's ability to remain still, sustain activity, or follow assessment procedures &
Clinicians provide rest breaks, adjust procedures, use play, and involve child life specialists and other support staff &
Design flexible assessment procedures and child-friendly interactions that can adapt to children's needs during data collection &
Support pediatric participation by allowing clinical technologies and procedures to adapt around the child rather than requiring the child to adapt to the technology \\

\midrule

CGA systems do not fully support the coordination and contextual knowledge required during pediatric assessment &
Child life specialists help children navigate assessments, while physical and occupational therapists provide additional knowledge about children's movement and behavior &
Design systems that support coordination and information sharing across clinicians, support staff, children, and families &
Recognize pediatric healthcare technologies as part of team-based care rather than interactions between a single clinician and patient \\

\midrule

Gait laboratories can affect children's comfort and movement, while a single clinical assessment provides only a snapshot of how they move &
Clinicians rely on observations, follow-up, and knowledge of children's movement outside the gait lab &
Support portable and longitudinal sensing across homes, schools, playgrounds, sports settings, and other everyday environments &
Design pediatric health technologies that can capture clinically relevant information across the environments where children live and move \\

\midrule

Current CGA can require substantial setup, operation, and data-processing work from clinical staff &
Clinicians and biomechanists manually configure systems, clean and process data, and coordinate information across the gait analysis workflow &
Simplify setup and operation and automate repetitive processing while maintaining clinically useful information &
Reduce the work required to use pediatric clinical technologies while considering how automation changes existing clinical roles \\

\bottomrule
\end{tabularx}
\end{table*}

\subsection{Limitations and Future Work}
Several limitations should be considered when interpreting these findings. First, our sample was drawn from a university-affiliated clinical network and may overrepresent well-resourced settings with access to specialized gait laboratories. Future work should include resource-constrained settings. Second, our modest sample included complementary clinical roles and a system designer, so the survey findings should be interpreted as descriptive expert perspectives rather than generalizable across professionals. Finally, our study relied on interviews, surveys, and scenario-based ideation rather than direct observation of gait assessments.

Our study also focused on clinician and designer perspectives and does not capture the experiences of children and their families. Future work should directly involve child patients and caregivers using age-appropriate and accessible methods to understand how they experience CGA and how sensing systems can better support their needs. Observational and prototype-based studies could also examine how sensor configurations and clinical environments and automation impact patient comfort, clinician workload, and data quality. 

\section{Conclusion}
Computerized gait analysis (CGA) can provide objective information for pediatric diagnosis and treatment planning, but its use in pediatric care depends on more than sensing accuracy and technical performance. Through a mixed-methods study with 12 pediatric clinicians and one system designer, we show how children's bodies, sensory needs, participation, clinical workflows, and care settings shape how CGA is used in practice. Participants valued the objective and information that CGA provides while also describing mismatches between the requirements of current systems and the needs of pediatric care. Clinicians and care teams often adapt sensing procedures to address these mismatches. Our findings point to design implications for CGA systems that better support pediatric patients, such as extending analysis across children's everyday environments.

\section*{Acknowledgments}
    This work was supported by the Stanford Graduate Fellowship and the Stanford Institute for Human-Centered Artificial Intelligence (HAI). We thank the clinicians who generously shared their time and expertise by participating in this study.

\bibliographystyle{ACM-Reference-Format}
\bibliography{references}


\clearpage
\onecolumn
\appendix

\section{Survey Instrument}
\label{app:survey}

Participants completed surveys tailored to their clinical role. The survey
consisted of the following components.

\vspace{0.75em}

\noindent\textbf{Medical Experts and Designers}

Participants completed three survey sections:

\begin{itemize}[leftmargin=*,itemsep=0.5em]

\item \textbf{Technology evaluation.}
Participants rated the importance of seven design criteria:
functionality, accuracy, ease of use, comfort, affordability,
accessibility, and scalability using a 7-point Likert scale
(1 = not important; 7 = extremely important).

\item \textbf{Sensor perception.}
Participants evaluated sensing technologies according to
functionality, accuracy, training requirements, patient comfort,
affordability, availability, and scalability using a three-level
scale (low, moderate, high).

\item \textbf{Context-specific deployment.}
Participants indicated whether each sensing technology should be used
in the gait laboratory, local care center, or the patient's home
using one of three options:
\emph{not use}, \emph{maybe use}, or \emph{definitely use}.

\end{itemize}

\vspace{1em}

\noindent\textbf{Physicians}

Participants completed three survey sections:

\begin{itemize}[leftmargin=*,itemsep=0.5em]

\item \textbf{Technology evaluation.}
Physicians rated the same seven technology evaluation criteria
(functionality, accuracy, ease of use, comfort, affordability,
accessibility, and scalability) using a 7-point Likert scale
(1 = not important; 7 = extremely important).

\item \textbf{Clinical decision-making.}
Physicians rated the importance of temporal parameters,
spatial parameters, joint angles, ground reaction forces,
and muscle activation using a three-level scale
(low, moderate, high).

\item \textbf{Context-specific gait parameters.}
Physicians rated the importance of each gait parameter for use in the
gait laboratory, local care center, and patient's home using a
three-level scale (low, moderate, high).

\end{itemize}

\end{document}